\documentclass{elsart}
\usepackage[utf8]{inputenc}
\usepackage[english]{babel}
\usepackage{graphicx}
\usepackage{amsmath}
\usepackage{amssymb}
\usepackage{cite}
\usepackage{enumitem}
\usepackage[draft]{hyperref}
\newcommand{\Bi}{$\mathrm{Bi}$}
\newcommand{\Biz}{$\mathrm{Bi}^{0}$}
\newcommand{\Bimi}{$\mathrm{Bi}^{-}$}

\newcommand{\Bipi}{$\mathrm{Bi}^{+}$}
\newcommand{\Bipii}{$\mathrm{Bi}^{2+}$}

\newcommand{\Bipv}{$\mathrm{Bi}^{5+}$}
\newcommand{\Biii}{$\mathrm{Bi}_2$}

\newcommand{\Biiim}{$\mathrm{Bi}_2^{-}$}
\newcommand{\Biiimm}{$\mathrm{Bi}_2^{2-}$}
\newcommand{\BiO}{$\mathrm{BiO}$}
\newcommand{\SiOiv}{\mbox{$\mathrm{SiO}_4$}}
\newcommand{\GeOiv}{\mbox{$\mathrm{GeO}_4$}}
\newcommand{\GeOvi}{\mbox{$\mathrm{GeO}_6$}}
\newcommand{\AlOiv}{\mbox{$\mathrm{AlO}_4$}}
\newcommand{\OPOiii}{\mbox{$\mathrm{O}\!=\!\mathrm{PO}_3$}}
\newcommand{\GeOii}{\mbox{$\mathrm{GeO}_2$}}

\newcommand{\gamess}{\mbox{GAMESS~(US)}}
\newcommand{\octopus}{\mbox{Octopus}}
\newcommand{\cmminusi}{$\textrm{cm}^{-1}$}
\begin{document}
\begin{frontmatter}
\renewcommand{\baselinestretch}{1}              % Single spacing in frontmatter
\title{%
Interstitial bismuth dimers and single atoms
as possible centres of broadband near-IR
luminescence in bismuth-doped glasses
}
\author{V.O.Sokolov\thanksref{*}},
\author{V.G.Plotnichenko},
\author{E.M.Dianov}
\address{Fiber~Optics~Research~Center of the~Russian~Academy~of~Sciences \\
38~Vavilov~Street,~~Moscow~~119333,~~Russia}
\thanks[*]{Corresponding author. \\
\mbox{Tel.:~+7~499~135~8093};
\mbox{Fax:~+7~499~135~8139};
\mbox{E-mail:~vence.s@gmail.com}, sokolov@fo.gpi.ac.ru}
\begin{abstract}
Absorption, luminescence and Raman spectra of interstitial bismuth atoms, \Biz,
and negatively charged dimers, \Biiim, in alumosilicate, germanosilicate,
phosphosilicate and phosphogermanate glasses networks are calculated by
time-dependent density functional method. On grounds of this calculation an
extension of our previously suggested model of broadband near-IR luminescence in
bismuth-doped glasses is put forward.
\end{abstract}
%\journal{Journal of Non-Crystalline Solids}
\end{frontmatter}

\section*{Introduction}
Near-IR broadband (1100 -- 1400~nm) luminescence in bismuth-doped glasses
discovered in Refs.~\cite{Murata99, Fujimoto01} is being studied intensively.
By now the luminescence has been observed in many bismuth-doped glasses, such as
alumosilicate (e.g. \cite{Murata99, Fujimoto01, Fujimoto06, Dianov06,
Khonthon07, Ren06, Denker09}), alumogermanate (e.g. \cite{Dianov06, Xia06,
Meng05c, Peng05c}), alumoborate (e.g. \cite{Peng05c, Meng05b, Denker07}),
alumophosphosilicate, alumophosphate, alumophosphoborate \cite{Dianov06,
Denker07, Razdobreev08, Meng05a}, chalcogenide \cite{Yang07, Hughes09} and in
several bismuth-doped crystals (RbPb$_2$Cl$_5$ \cite{Butvina08}, FAU-type
zeolites \cite{Sun09}, BaB$_2$O$_4$ \cite{Su09}). The bismuth-related IR
luminescence is used successfully in laser amplification and generation (see,
e.g. \cite{Dianov07, Dianov08a}). The present state of research and applications
of near-IR luminescence in bismuth-doped glasses are examined in the recent
reviews, Ref.~\cite{Dianov08b, Dianov09}.

However there is no commonly accepted model of the IR luminescence center.
Several models are suggested, such as electronic transitions in \Bipi{}
\cite{Meng05b, Meng05a, Razdobreev08, Yang07}, \Bipii{} \cite{Yang07} and
\Bipv{} \cite{Fujimoto01, Fujimoto06, Dianov06, Xia06} interstitial ions, in
\BiO{} interstitial molecules \cite{Ren06}, \Biii, \Biiim{} and \Biiimm{}
interstitial dimers \cite{Khonthon07, Meng05c, Sokolov08, Sokolov09, Hughes09},
other bismuth  clusters \cite{Meng05c}, \mbox{$\mathrm{BiO}_4$} complexes with
tetrahedral coordination of the central bismuth ion \cite{Kustov09}.

In our opinion there is good evidence for negative bismuth dimers being the
luminescence centres (see e.g. \cite{Sokolov09, Denker09,Hughes09}). However,
up to now, there are no convincing explanation of changes in absorption and
luminescence spectra with glass composition (see e.g. Refs.~\cite{Dianov09,
Dianov08a}). In the context of the \Biiim{} and \Biiimm{} dimers as centres of
the IR luminescence it would appear reasonable that such changes are caused by
Stark shift and splitting together with certain rearrangement of dimer states in
glass network electric field. The latter does obviously depend on glass
composition.

Alternatively, there is an evidence that single Bi atoms and ions do occur as
well in the bismuth-doped glass host. We studied Raman scattering of Ar laser
457.9~nm-light in optical fibres with bismuth-doped alumosilicate glass core and
observed a narrow intensive band in the low-frequency part of the Raman spectra,
near 110~\cmminusi{} (Ref.~\cite{Sokolov09}, Fig.~4, ''a'' band). Our
calculation \cite{Sokolov09} of neutral single bismuth atom, \Biz, and negative
single bismuth ion, \Bimi, incorporated in alumosilicate host in the sixfold
ring interstitial sites proved both \Biz{} and \Bimi{} have three vibrational
modes practically not mixed with any vibrations of the rings. The frequency of
these vibrational modes are found to fall in the 75 -- 105 and 85 --
130~\cmminusi{} ranges, for \Bimi{} and \Biz, respectively (see Fig.~5 in
Ref.~\cite{Sokolov09}). One mode corresponds to displacement of the interstitial
bismuth atom along the rings axis and the other two correspond to transverse
displacement of the atom. All three vibrational modes are Raman active,
scattering in the transverse vibrations being an order as intensive as in the
longitudinal ones. Hence the Raman band near 110~\cmminusi{} in bismuth-doped
alumosilicate glass may be indicative of single bismuth atoms, \Biz, or negative
ions, \Bimi{} in sixfold ring interstitial sites of the glass network. Recently
electronic transitions in the \Biz{} atom has been supposed to be responsible
for the near-IR luminescence in bismuth-doped glasses \cite{Peng09}.

The aim of this work were to verify the assumptions regarding both the network
electric field and the interstitial bismuth atoms by immediate calculation of
optical properties of interstitial Bi atoms and ions in networks of silica- and
germania-based glasses.

\section*{Calculations}
In our previous works \cite{Sokolov08, Sokolov09} we used calculated data for
free negatively charged \Biii{} dimers to explain absorption and luminescence
spectra of bismuth-doped glass. Experimental data available for free \Biz{}
atom were used to the same purposes in Ref.~\cite{Peng09}. However an influence
of glass network (interstitial electrostatic field, in essential) must be taken
into account to discuss optical spectra of interstitial dimers and atoms.

We performed quantum-chemical calculation of interstitial electric field in
network of glasses of several compositions (silica, germania, alumosilicate,
germanosilicate, phosphosilicate, phosphogermanate). Cluster approach was used
to model the atomic environment of interstitial sites. Two coaxial sixfold
rings each formed by \SiOiv, \GeOiv{} (for germanosilicate glasses with low
germania content), \AlOiv{} or \OPOiii{} tetrahedra or \GeOvi{} octahedra (for
germanosilicate glasses with high germania content) were incorporated in proper
numbers in the clusters. The distance between the ring centres was optimized
during the calculations. To ensure aluminum atoms being fourfold coordinated,
extra electrons were placed in the aluminum-containing clusters, one electron
per aluminum atom. In \GeOvi-containing clusters two extra electrons were added
per sixfold coordinated germanium atom to make it be stable in this
coordination. Dangling bonds of the outer oxygen atoms in the clusters were
saturated with hydrogen atoms. All calculations were carried out with \gamess{}
quantum-chemical code \cite{gamess} by DFT method using BLYP functional which is
known to provide nice agreement of calculated geometrical parameters with
experimental data. We used the bases and effective core potentials developed in
Ref.~\cite{SBKJC}. One $d$-type polarization function with $\zeta =
0.8000$~a.u.{} was added in the basis for each oxygen atom. Standard
\mbox{3-21G} basis was used for hydrogen atoms.

\begin{figure}[h]
\begin{center}
\includegraphics[scale=0.61,bb=15 210 660 780]{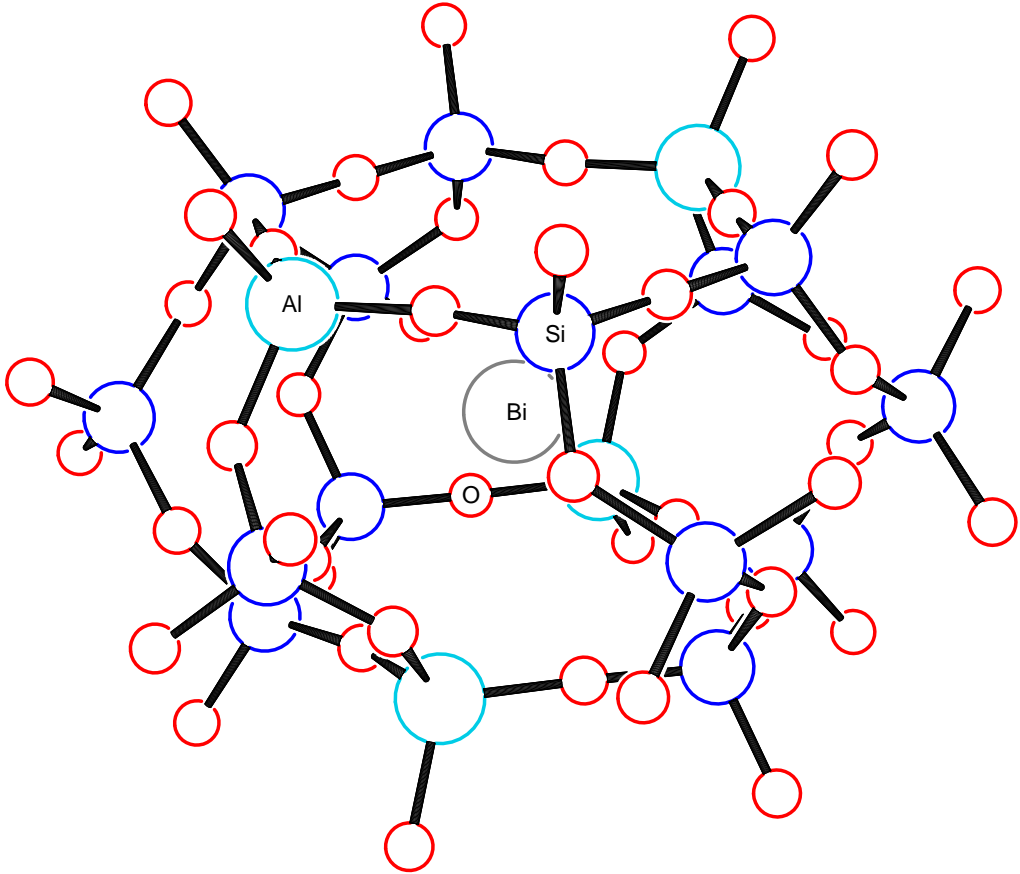}
\end{center}
\caption{Calculated configuration of \Biz{} atom in the sixfold-rings
interstitial site of alumosilicate glass}
\label{fig:1}
\end{figure}

Quantum-chemical modelling of configuration and vibrational properties of
\Biii{}, \Biiim{} and \Biiimm{} dimers and of single \Biz{} atom and \Bimi{} ion
in alumosilicate network was performed in our previous works \cite{Sokolov08,
Sokolov09} using the above-described cluster approach. Calculated configuration
for \Biiim{} interstitial dimer was presented in figure~3 of
Ref.~\cite{Sokolov09} for alumosilicate glass. In the present work such
calculations were repeated for silica, germania, germanosilicate,
phosphosilicate, and phosphogermanate clusters. The equilibrium configurations
of negatively charged dimers, both \Biiim{} and \Biiimm, in the interstitial
site formed by two sixfold rings are found to be quite similar in all these
networks: the dimers are aligned along the common axis of the rings, the first
bismuth atom being practically in the center of one ring and the second atom
being between two rings. Again, equilibrium position of \Biz{} atom and \Bimi{}
ion are found to be between two six-rings in all these networks. As a
representative example, calculated configuration of alumosilicate glass cluster
with \Biz{} interstitial atom is shown in figure~\ref{fig:1}.
Such configurations turns out to be highly stable for both dimers and atom or
ion: bismuth atoms do not form bond with any atom of the rings and returned
to the equilibrium positions even after dimer or single atom is displaced
considerably from that position. Vibrational properties of interstitial \Biz{}
atom and \Biiim{} dimer in all these networks are found to be quite similar to
those calculated in Ref.~\cite{Sokolov09} for alumosilicate host (see above).

Electric field strength was calculated in the point of the \Biz{} atom position
(or, what is the same, near the bismuth atom of the \Biii{} dimer which is
situated between two sixfold rings) and in several (8 -- 10) points in its
vicinity. The value averaged over all such points was used in further
calculations of optical spectra. The mean electric field strengths in ring
interstitial sites calculated for several glass compositions are given in
table~\ref{tab:E-fields}. As evident from these data, the interstitial electric
field is found to be approximately the same in silica, germanate and
germanosilicate (with low germania content) hosts but considerably higher both
in glasses containing aluminate or phosphate component and in germanosilicate
glasses with high germania content. This is attributable to charged (\AlOiv{}
and \GeOvi) or strongly dipole-polarized (\OPOiii) structural units in the rings
around interstitial bismuth sites in such glasses.
\begin{table}[h]
\caption{Calculated electric field in sixfold ring interstitial sites of glass
network}
\vspace{\baselineskip}
\begin{center}
\begin{tabular}{l|c}
\hline\hline & \\ [-1.5ex]
Ring composition  & Electric field, 0.001~a.u. \\ [0.5ex]
\hline\hline & \\ [-1.5ex]
6\SiOiv           &  2 \\
6\GeOiv           &  2 \\
4\SiOiv\,--\,2\GeOiv  &  2 \\
4\SiOiv\,--\,2\AlOiv  &  5 \\
4\SiOiv\,--\,2\GeOvi  &  7 \\
4\SiOiv\,--\,2\OPOiii &  7 \\
4\GeOiv\,--\,2\OPOiii & 10 \\ [0.5ex]
\hline\hline
\end{tabular}
\end{center}
\label{tab:E-fields}
\end{table}

Optical spectra calculations were performed by time-dependent density functional
theory (TDDFT) method using Octopus program \cite{octopus} and
Hartwigsen-Goedecker-Hutter pseudopotentials \cite{Hartwigsen98} with spin
polarization and spin-orbit interaction taken into account. PBE density
functional \cite{PBE} was used in the ground-state calculation. To obtain the
linear optical absorption spectrum of the system, the Octopus code excites all
frequencies of the system by giving certain (small enough) momentum to the
electrons and then evolves the time-dependent Kohn-Sham equations in real space
for a certain real time \cite{Bertsch00}. The dipole-strength function (or the
photo-absorption cross section) is then obtained by a Fourier transform of the
time-dependent dipole moment. Adiabatic LDA approximation is used in these
calculations to describe exchange-correlation effects. The Octopus code uses
real-space uniform grid inside the sum of spheres around each atom of the system
(a single one in our case). The sphere radius and the grid spacing were taken to
be was 8.0 and 0.25~\AA, respectively, in our calculations. The real-time
propagation was performed with $2\cdot 10^4$ time steps with the total
simulation time of about 20~fs. The Fourier transform was performed using
third-order polynomial damping (see \cite{octopus} for details of the code).

\begin{figure}[h]
\begin{center}
\includegraphics[scale=0.81,bb=60 290 540 780]{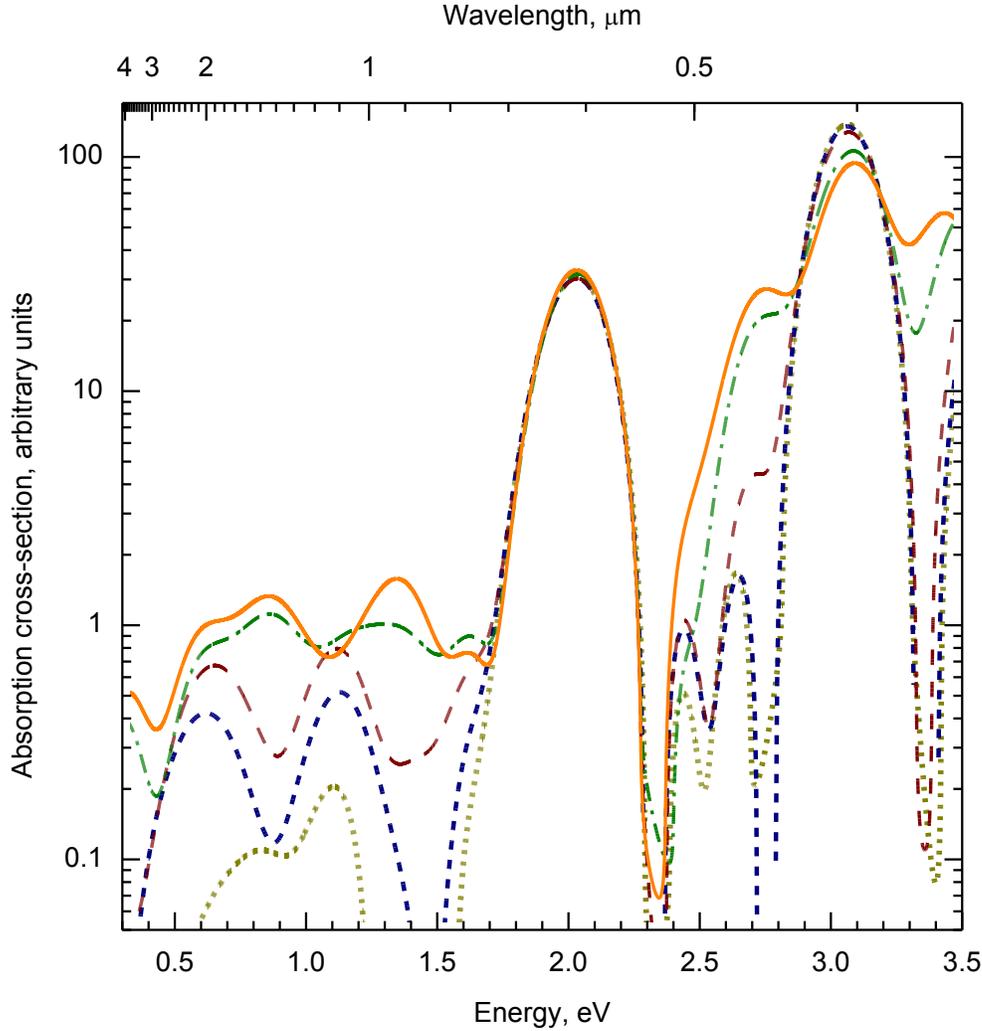}
\end{center}
\caption{Optical cross-section spectra of free \Biiim{} dimer (dots) and of
interstitial \Biiim{} dimers in silica, germania or germanosilicate (with low
germania content) host (short dashes), alumosilicate host (dashes), in
phosphosilicate or germanosilicate (with high germania content) host (dashes
plus dots) and in phosphogermanate host (solid)}
\label{fig:2}
\end{figure}

\begin{figure}[h]
\begin{center}
\includegraphics[scale=0.81,bb=60 290 540 780]{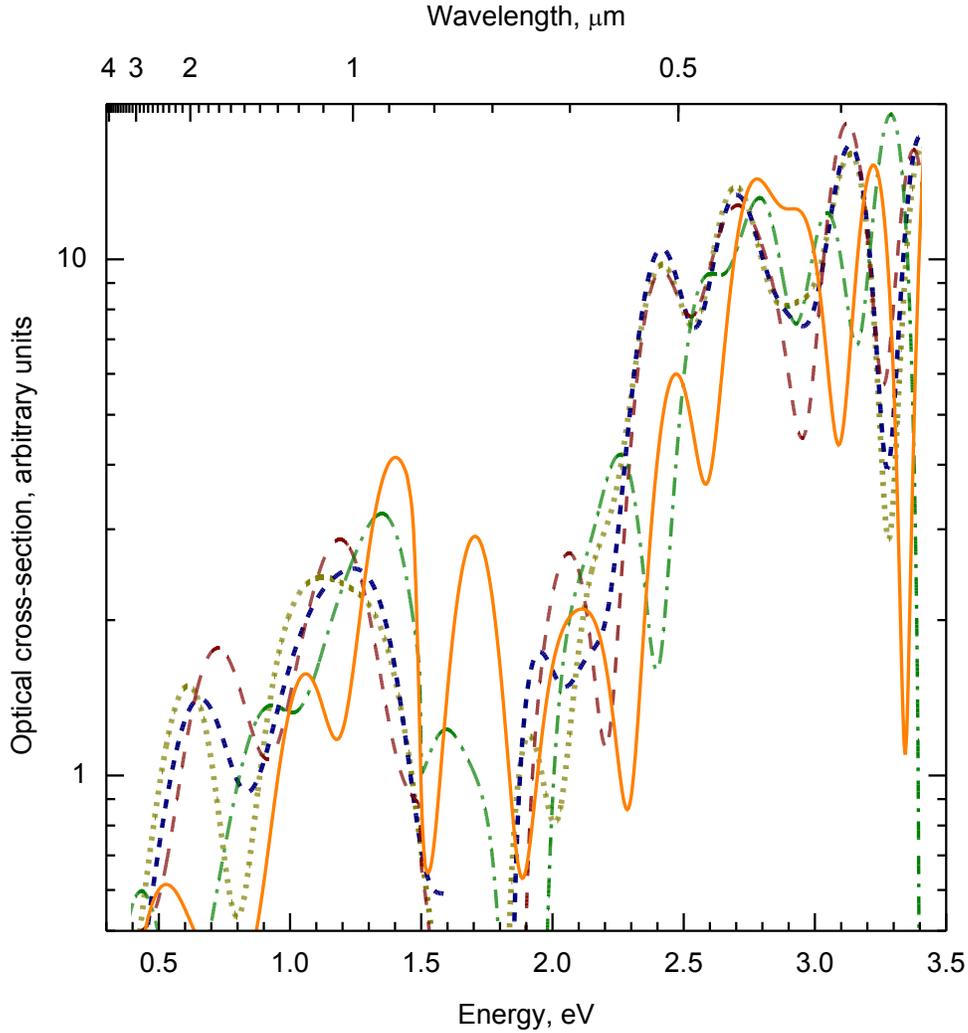}
\end{center}
\caption{Optical cross-section spectra of free \Biz{} atom (dots; the part
for energy $< 1.6$~eV multiplied by 10 for clarity) and of interstitial \Biz{}
atom in silica, germania or germanosilicate  (with low germania content) host
(short dashes), alumosilicate host (dashes), in phosphosilicate or
germanosilicate (with high germania content) host (dashes plus dots) and in
phosphogermanate host (solid)}
\label{fig:3}
\end{figure}

\begin{figure}[h]
\begin{center}
\includegraphics[scale=0.81,bb=60 290 540 780]{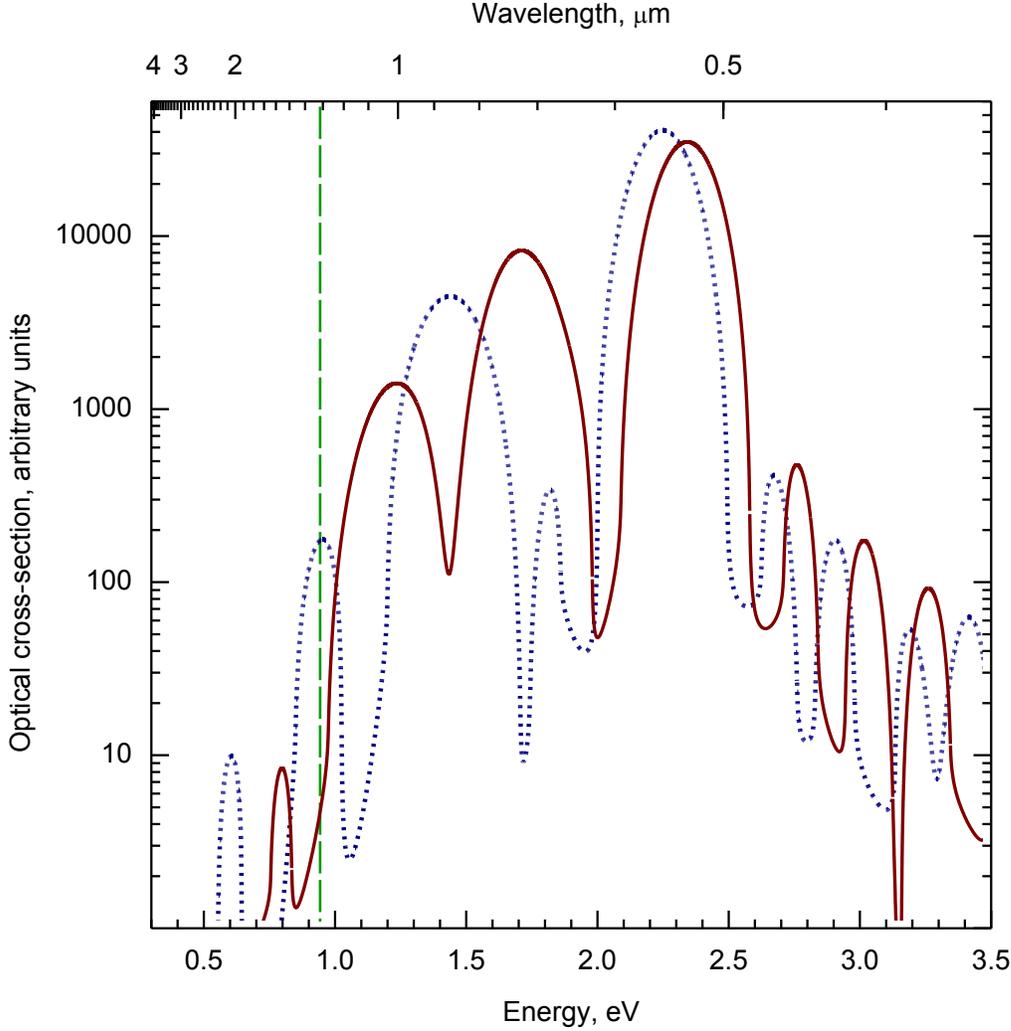}
\end{center}
\caption{Optical cross-section spectra of free \Bimi{} ion (dotted line) and of
interstitial \Bimi{} ion in alumosilicate host (solid line).}
\label{fig:4}
\end{figure}

\begin{figure}[h]
\begin{center}
\includegraphics[scale=0.50,bb=-10 -20 620 1010]{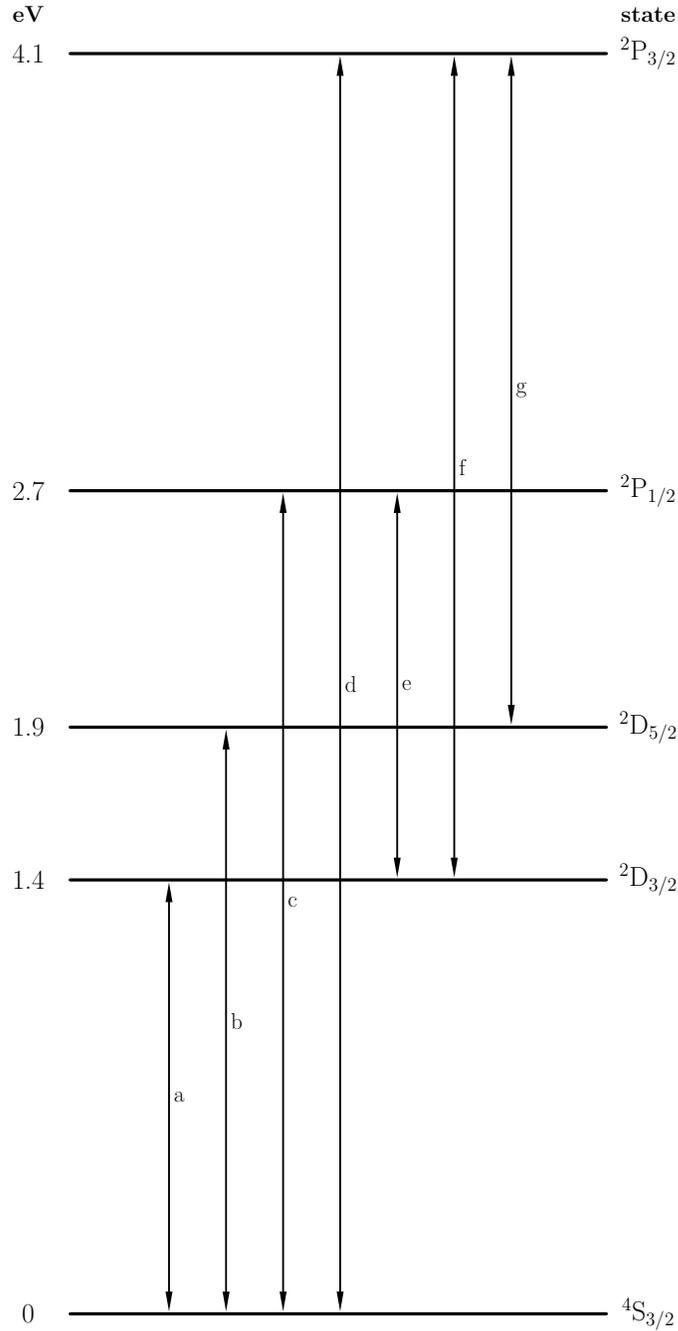}
\end{center}
\caption{States, transition types and lifetimes in \Biz{} atom corresponding to
the $6\mathrm{s}^2 6\mathrm{p}^3$ configuration \cite{Garstang64, George85,
Kozlov96, Bilodeau01}:
(a)~876~nm, M1, 32~ms;
(b)~648~nm, M1, 156~ms;
(c)~462~nm, M1, 18~ms and E2, 161~ms;
(d)~302~nm, M1, 137~ms;
(e)~976~nm, M1, 833~ms and E2, 222~ms;
(f)~460~nm, M1, 8~ms;
(g)~564~nm, M1, 43~ms}
\label{fig:5}
\end{figure}

Shown in figure~\ref{fig:2} are the calculated optical spectra of free
\Biiim{} dimer and of interstitial \Biiim{} dimers in sixfold ring
interstitial sites of networks of silica (or germania or germanosilicate),
alumosilicate, phosphosilicate and phosphogermanate glasses.
Figures~\ref{fig:3} and \ref{fig:4} display similar spectra for free
and interstitial \Biz{} atoms and for free and interstitial \Bimi{} ion,
respectively. Compairing the results for free \Biiim{} dimers with previous
configuration interaction calculations \cite{Balasubramanian91, Sokolov08,
Sokolov09} and those for free \Biz{} with experimental and theoretical data
available \cite{Garstang64, George85, Kozlov96, Bilodeau01} (see
figure~\ref{fig:5}), one readily sees that both transition wavelengths
and transition intensities are described adequately in our TDDFT calculations.

Figure~\ref{fig:2} shows calculated optical spectra of interstitial
\Biiim{} dimer in comparison with those of a free dimer. Similar spectra for
interstitial \Biz{} atom are presented in figure~\ref{fig:3}.
From figure~\ref{fig:2} it will be noticed how optical spectrum of an
interstitial \Biiim{} dimer is modified in comparison with free dimer.
According to our previous calculations performed by configuration interaction
(CI) method \cite{Sokolov09}, absorption at 860, 720, 460~nm and $\lesssim
400$~nm wavelengths in the free \Biiim{} dimer is caused by transitions from
the ground state to the excited ones. The IR luminescence at 1450, 1300 and
1050~nm wavelengths corresponds to spin-forbidden transitions from three lowest
excited states to the ground state and the visible luminescence at 750~nm is
caused to a spin-allowed transition with low oscillator strength from one of
the other excited states to the ground state. Notice that compairing results of
CI calculations with optical spectra calculated in TDDFT with \octopus{}
package (such as shown in figures~\ref{fig:2}, \ref{fig:3}, \ref{fig:4}) one
should realize that the latters include all possible transitions in the system
under consideration.

For the interstitial \Biiim{} dimers, the absorption in the $\lesssim 400$~nm
range and transitions corresponding to the absorption and visible luminescence
in the 600 -- 750~nm is found in our calculations to be practically unaltered in
all the surrounding glass network compositions studied. The absorption band near
500~nm is found to be shifted towards shorter wavelengths (near 440~nm in
phosphogermanate environment) and somewhat intensified. Besides, absorption near
800~nm growths. In phosphate-bearing and in germanosilicate (with high \GeOii{}
content) glasses absorption in the 900 -- 1000~nm range becomes significant. In
glasses with no phosphate component present and in germanosilicate glasses with
low \GeOii{} content, the IR luminescence is not too different from that in free
dimers and occurring mainly in the 1000 -- 1300~nm range. When passing to
phosphate-bearing compositions and for high \GeOii{} content one finds the IR
luminescence bands to be shifted towards longer wavelengths, in the 1400 --
1600~nm range.

Again, optical spectra of \Biz{} interstitial atoms are found in our
calculations to change compared with those of free atoms
(figure~\ref{fig:3}). This change is relatively low in alumosilicate glasses
where \Biz{} interstitial atoms give rise to absorption bands at $\lesssim
400$~nm, near 500~nm and in 600 -- 700 and 900 -- 1100~nm ranges. This
absorption causes IR luminescence in 1000 -- 1300 and 1400 -- 1600~nm wavelength
ranges. However the change turn out to be much more pronounced in
phosphosilicate glasses and in germanosilicate glass with high germania content
where the absorption should occur in 500 -- 600~nm wavelength range and,
especially, in 700 -- 1000~nm range. The corresponding IR luminescence in these
glasses is found to arise in 1000 -- 1400~nm range and near 2000~nm.

The above-described modification of optical spectra of bismuth interstitial
dimers and atoms with composition of surrounding glass network gives an insight
into origin of variations of absorption and IR luminescence bands discovered in
bismuth-doped glasses \cite{Dianov08a, Dianov09}. In particular, characteristic
IR luminescence band shift from $\sim 1100$~nm into the 1200 -- 1400~nm range is
observed while turning from alumosilicate compositions to phosphate-bearing
ones.

Ionization potential of the negative bismuth ion, \Bimi, is known to be rather
low: e.g. in Ref.~\cite{Bilodeau01} the electron affinity of ${}^{209}$Bi is
found to be $7600\textrm{~cm}^{-1}$ or $0.94$~eV. Because of this, all
transitions in $\lesssim 1300$~nm wavelength range should, in general, result
in \Bimi{} ionization with neutral \Biz{} atom formed and free electron occurred
in conduction band. Since optical cross-sections in \Bimi{} ion are at least an
order as high as those in \Biz{} atom, one would expect the interstitial
\Bimi{} ions are destroyed rapidly under both visible and IR irradiation. Thus
interstitial \Bimi{} ions in glass may promote formation of interstitial
negatively charged \Biii{} dimers \cite{Sokolov09}. On the other
hand, recombination luminescence in 1500 -- 1700~nm range is not improbable
owing to conduction electrons capturing in \Biz{} interstitial atoms with
\Bimi{} ions formed.

\section*{Discussion}
We have established, on modelling grounds, that interstitial negatively charged
bismuth dimers, \Biiim{} and \Biiimm, can actually occur in glass network and
that absorption and luminescence spectra of the dimers correlate well with the
experimental spectra of bismuth-doped glasses. On the other hand, bismuth
interstitial atoms, \Biz, and negative ions, \Bimi, are demonstrated to occur
in glass network as well. In particular, certain features observed in Raman
spectra of bismuth-doped alumosilicate glasses are attributable to both \Biii{}
dimers and \Bi{} atoms (ions) present in glass together.

So it would appear reasonable that both \Biz{} interstitial atoms and \Biiim{}
(\Biiimm) negatively charged dimers occur in  in bismuth-doped glasses as two
types of the IR luminescence centres. Relation between interstitial \Biii{}
dimers and \Bi{} atoms (and ions) concentrations in the given glass is likely to
depend predominantly on bismuth dopant content. It should be stressed that in
bismuth-doped glasses considerable part of the dopant bismuth atoms are likely
to be bonded with oxygen atoms of surrounding network. However such bismuth
centres bear no relation to IR luminescence.

It follows from the relation between calculated intensities of transitions
corresponding to absorption and IR luminescence in \Biz{} atom that the IR
luminescence time constant is considerably less than in \Biiim{} or \Biiimm{}
dimers and that it decreases rapidly with electrostatic field growth
(figure~\ref{fig:3}). Similar calculations of \Biii{} dimers in electrostatic
field prove the transition intensities to change only slightly and the
transition wavelengths to change almost as rapidly as in \Biz{} atom with the
field strength. Therefore, interstitial \Biz{} atoms should give rise to much
faster IR luminescence in comparison with \Biiim{} and \Biiimm{} dimers.
Unfortunately the transition intensities in atoms and dimers are hardly to be
compared directly in our calculational approach. Rough estimate results in ratio
of the order of 10. Thus, interstitial \Biz{} atoms may be responsible for
''fast'' component observed in IR luminescence.

The calculations described in the present article and in Ref.~\cite{Sokolov08,
Sokolov09} are concerned with sixfold ring interstitial sites. The sixfold
ring are known to be the most abundant ring structure in silica- and
germania-based glasses network. However there are other rings as well in such
glasses, mainly five-, seven- and eightfold ones. We have performed some
calculations for interstitial sites formed by pairs of such rings in
alumosilicate network to examine equilibrium states of bismuth atoms and dimers
and electric field strength. It is felt that the results and conclusions
obtained for the sixfold rings still remain valid for sevenfold and,
supposedly, in eightfold rings. The fivefold rings turn out to be too
narrow to hold the interstitial dimers and, supposedly, atoms.

\section*{Conclusion}
In summary, we calculated optical spectra of interstitial negatively charged
bismuth dimers, \Biiim, neutral atoms, \Biz, and negative ions, \Bimi, in
network of silica- and germania-based glasses with various composition.
Considering our previous works \cite{Sokolov08, Sokolov09}, the results of the
present calculations allow definite conclusions, as follows.
\renewcommand{\labelitemi}{---}
\begin{itemize}
\item Both interstitial \Biiim{} and \Biiimm{} dimers and interstitial \Biz{}
atoms and \Bimi{} ions can occur in bismuth-doped silica- and germania-based
glasses. IR luminescence in 1000 -- 1600~nm in these glasses can be caused by
all such interstitial centres.
\item IR luminescence decay time in \Biz{} atoms is estimated to be an order as
short as in \Biiim{} or \Biiimm{} dimers . Hence \Biiim{} (\Biiimm) dimers and
\Biz{} atoms can be assumed to cause ''slow'' and ''fast'' parts of the IR
luminescence, respectively.
\item Interstitial electric field strength dependence of absorption and IR
luminescence in the interstitial \Biiim{} (\Biiimm) dimers is found to differ
considerably from that in the interstitial \Biz{} atoms. Pronounced dependence
of the IR luminescence spectra and its excitation spectra on glass composition
and bismuth content observed in several experiments may be attributable to the
interstitial field variations.
\item interstitial \Biz{} atom is found to capture under certain conditions an
electron from the glass conduction band with \Bimi{} formed. Recombination
luminescence in the 1500 -- 1700~nm is likely to occur owing to such capturing.
On the other hand, the interstitial negative \Bimi{} ions are readily destroyed
under $\lesssim 1300$~nm irradiation.
\end{itemize}

Thus our results of the present work and of Ref.~\cite{Sokolov08, Sokolov09}
concerning interstitial bismuth atoms and negatively charged bismuth dimers in
bismuth-doped glasses provide support for the conclusion that both \Biz{} and
\Biiim{} (\Biiimm) occur in these glasses as two types of the IR luminescence
centres.

\ack
We are grateful to Prof.~I.A.Bufetov for valuable discussions.

\newpage
\pagestyle{empty}
%
% References %%%%%%%%%%%%%%%%%%%%%%%%%%%%%%%%%%%%%%%%%%%%%%%%%%%%%%%%%%%%%%%%%%%
%
\renewcommand{\baselinestretch}{1}               % Single spacing
\end{document}